# Georeactor Variability and Integrity

by

J. Marvin Herndon

Transdyne Corporation
San Diego, California 92131 USA

October 3, 2005

Communications:
mherndon@san.rr.com          http://NuclearPlanet.com          http://UnderstandEarth.com

Applying Fermi's nuclear reactor theory, I demonstrated the feasibility of planetocentric nuclear fission reactors, which offer possible explanations for the internal energy produced by Jupiter, Saturn, and Neptune (Herndon 1992), for the energy source that powers the geomagnetic field (Herndon 1993; Herndon 1994), and, perhaps, for energy production associated with other planets and large moons in our Solar System.

As a deep-Earth energy source, the planetocentric nuclear-fission georeactor concept is on a more secure scientific footing than the previous idea related to the assumed growth of the inner core. There is no observational support for the idea that the inner core is cooling and growing (Herndon 2005), and no legitimate basis for believing that useful energy, concomitantly produced, would power the geomagnetic field instead of simply reducing the inner-core rate of cooling. In striking contrast, the close agreement between observed deep-Earth helium isotope ratios and georeactor-produced helium isotopes (Herndon 2003), calculated from state-of-the-art numerical simulations (Hollenbach & Herndon 2001), is strong evidence for the existence of the nuclear georeactor. Yet, despite that, some would – by omission – mislead the broad, general-interest, scientific community (Araki et al. 2005; McDonough 2005). At the foundations of science, science and ethical considerations are inseparable.

Previously envisioned deep-Earth energy sources, including natural radioactivity, have one characteristic in common: Change is possible only gradually and in only one direction over time: On a human time-scale, that is essentially no change at all. Recently, Fogli et al. (2005) performed a likelihood analysis of the neutrino event spectra in energy and time using monthly-binned data on event-by-event energies in KamLAND and on power reactors in Japan. In making their calculations, Fogli et al. (2005) assumed that georeactor energy output is constant. Within the framework of present understanding, the assumption of constant georeactor output may or may not be valid.

The purpose of this brief communication is to emphasize the importance of scientific integrity and to highlight the possibility of variable georeactor power output so that these might be borne in mind in future investigations. Generally, variability in nuclear fission reactors arises from changes in composition and/or position of fuel, moderators, and neutron absorbers. Although as yet there is no irrefutable evidence of planetocentric nuclear reactor variability, circumstantial evidence certainly invites inquiry.

In the late 1960s, astronomers discovered that Jupiter radiates into space approximately twice as much energy as it receives from the Sun. Later, Saturn and Neptune were also found to radiate prodigious quantities of internally generated energy. Stevenson (1978), discussing Jupiter, stated, "The implied energy source ... is apparently gravitational in origin, since all other proposed sources (for example, radio-activity, accretion, thermonuclear fusion) fall short by at least two orders of magnitude…". Similarly, Hubbard (1990) asserted, "Therefore, by elimination, only one process could be responsible for the luminosities of Jupiter, Saturn, and Neptune. Energy is liberated when mass in a gravitationally bound object sinks closer to the center of attraction ... potential energy becomes kinetic energy …."

Having knowledge of the natural reactors at Oklo, I realized a different possibility and proposed the idea of planetary-scale, planetocentric nuclear fission reactors as energy sources for the giant planets (Herndon 1992), demonstrating their feasibility in part using the same calculations employed in the design of commercial nuclear reactors and employed by Kuroda (1956) to predict conditions for the natural reactors that were discovered in 1972 at Oklo, Republic of Gabon.

The near-surface natural reactors at Oklo, which were critical about 1.8 billion years ago, operated intermittently (Maurette 1976). Recent investigations suggest quite rapid cycling periods with 0.5 hour of operation followed by 2.5 hours of dormancy (Meshik et al. 2004). While the specific control mechanism, presumably involving water, may not be directly applicable to planetocentric reactors, the observations nevertheless demonstrate the potential variability of natural nuclear reactors.

Atmospheric turbulence in the giant planets appears to be driven by their internal energy sources. Jupiter, Saturn, and Neptune produce prodigious amounts of energy and display prominent turbulent atmospheric features. Uranus, on the other hand, radiates little, if any, internally generated energy and appears featureless.

In the summer of 1878, Jupiter's Great Red Spot increased to a prominence never before recorded and, late in 1882, its prominence, darkness, and general visibility began declining so steadily that by 1890 astronomers thought that the Great Red Spot was doomed to extinction. Changes have been observed in other Jovian features, including the formation of a new lateral belt of atmospheric turbulence (Peek 1958).

Jupiter, 98% of which consists of a mixture of H and He, excellent heat transfer media, is capable of rapid thermal transport. It is important to establish whether these atmospheric changes are due to changes in planetocentric nuclear reactor output as it seems, especially as these would represent short-period variability (Herndon 1994).

Investigations of terrestrial heat flux developed under the assumption of essentially constant heat flow produced solely by long-lived, natural radioactive decay. That assumption should be questioned and efforts should be made to determine whether variations can be resolved.

Deep within the Earth, the geomagnetic field varies in intensity and reverses polarity frequently, but quite irregularly, with an average time between reversals of about 200,000 years. I have suggested that the variable and intermittent changes in the intensity and direction of the geomagnetic field may have their origin in nuclear reactor variability (Herndon 1993), but that has not yet been established. Ultimately, the nature and possible variability of deep-Earth energy production can be revealed by making fundamental discoveries and by discovering fundamental quantitative relationships in nature (Herndon 2005); not by making blatant misrepresentations (Araki et al. 2005; McDonough 2005).